\documentclass[amsmath,amssymb,aps,pra,superscriptaddress,floats,twocolumn,10pt]{revtex4-1} 

\usepackage{float}
\fontfamily{ptm}\selectfont
\usepackage{stmaryrd}
\usepackage[usenames,dvipsnames]{color}
\usepackage{graphicx,subfigure}
\usepackage{wrapfig}
\usepackage{xcolor}
\usepackage{cancel}
\usepackage{amsmath,amssymb}
\usepackage{times}
\usepackage[latin1]{inputenc}
\usepackage{setspace} 
\usepackage[normalem]{ulem}
\usepackage[hypertexnames=false,plainpages=false,pdfpagelabels,colorlinks=true,linkcolor=blue,urlcolor=blue,citecolor=blue,pdfdisplaydoctitle=true,pdfpagelayout=OneColumn,pdfduplex=DuplexFlipLongEdge]{hyperref}
\usepackage{textcomp}
\usepackage{multirow}
\usepackage{bm}
\usepackage{mathrsfs}

\newcommand{\ket}[1]{\ensuremath{{\vert #1 \rangle}}}
\newcommand{\bra}[1]{\ensuremath{\langle #1 \vert}}
\renewcommand{\vec}[1]{\ensuremath{\mathbf{#1}}}
\newcommand{\uvec}[1]{\ensuremath{\mathbf{\hat{#1}}}}
\newcommand{\abs}[1]{\ensuremath{\left\vert #1 \right\vert}}

\newcommand{\Q}{\ensuremath{\vec{Q}}}
\newcommand{\dphi}{\ensuremath{\varphi}}
\newcommand{\phimw}{\ensuremath{\varphi_{MW}}}

\newcommand{\nup}{\ensuremath{n_\uparrow}}
\newcommand{\ndown}{\ensuremath{n_\downarrow}}

\newcommand{\upm}{\ensuremath{u^{\pm}_{\text{ \textbf{\emph{k}}}}}}
\newcommand{\p}{\ensuremath{\mathbf{p}}}

\renewcommand{\aa}{\ensuremath{\mathbf{a}}}
\newcommand{\rr}{\ensuremath{\mathbf{r}}}
\newcommand{\Rr}{\ensuremath{\mathbf{R}}}

\newcommand{\tpsi}{\ensuremath{\tilde{\Psi}}}
\newcommand{\fs}{\ensuremath{\bm{\mathcal{F}}_\mu}}
\newcommand{\f}[1]{\ensuremath{\bm{\mathcal{F}}_{#1}}}
\newcommand{\delk}{\ensuremath{\nabla_\vec{k}}}
\renewcommand{\k}{\ensuremath{\vec{k}}}
\newcommand{\phidyn}{\ensuremath{\varphi_{\mathrm{dyn}}}}
\newcommand{\phig}{\ensuremath{\varphi_\mathrm{Berry}}}
\newcommand{\Htb}{\ensuremath{H_\mathrm{tb}}}
\newcommand{\KK}{\ensuremath{\vec{K}}}
\newcommand{\KKp}{\ensuremath{\vec{K'}}}
\newcommand{\dd}{\ensuremath{\vec{d}}}
\newcommand{\ww}{\ensuremath{\delta k_\Omega}}
\newcommand{\kyfin}{\ensuremath{k_y^\mathrm{fin}}}

\usepackage{ulem}

\newcommand{\editout}[1]{{\bf \sout{#1}}}
\renewcommand{\editout}[1]{}

\begin{document}

\singlespacing
\noindent
\title{An Aharonov-Bohm interferometer for determining Bloch band topology}

\newcommand{\lmu}{Fakult\"{a}t f\"{u}r Physik, Ludwig-Maximilians-Universit\"{a}t M\"{u}nchen, Schellingstr.\ 4, 80799 Munich, Germany}
\newcommand{\mpq}{Max-Planck-Institut f\"{u}r Quantenoptik, Hans-Kopfermann-Str.\ 1, 85748 Garching, Germany}
\newcommand{\stanford}{Department of Physics, Stanford University, Stanford, California 94305, USA}
\newcommand{\mail}{E-mail: }

\noindent
\author{L.\ Duca}
\affiliation{\lmu}
\affiliation{\mpq}

\author{T.\ Li}
\affiliation{\lmu}
\affiliation{\mpq}

\author{M.\ Reitter}
\affiliation{\lmu}
\affiliation{\mpq}

\author{I.\ Bloch}
\affiliation{\lmu}
\affiliation{\mpq}

\author{M.\ Schleier-Smith}
\affiliation{\stanford}

\author{U.\ Schneider}
\affiliation{\lmu}
\affiliation{\mpq}


\begin{abstract}
The geometric structure of an energy band in a solid is fundamental for a wide range of many-body phenomena in condensed matter and is uniquely characterized by the distribution of Berry curvature over the Brillouin zone. In analogy to an Aharonov-Bohm interferometer that measures the magnetic flux penetrating a given area in real space, we realize an atomic interferometer to measure Berry flux in momentum space. We demonstrate the interferometer for a graphene-type hexagonal lattice, where it has allowed us to directly detect the singular $\pi$ Berry flux localized at each Dirac point. We show that the interferometer enables one to determine the distribution of Berry curvature with high momentum resolution. Our work forms the basis for a general framework to fully characterize topological band structures and can also facilitate holonomic quantum computing through controlled exploitation of the geometry of Hilbert space.

\end{abstract}

\maketitle

More than thirty years ago, Berry \cite{Berry1984} delineated the effects of the geometric structure of Hilbert space on the adiabatic evolution of quantum mechanical systems. These ideas have found widespread applications in science \cite{Wilczek89} and are routinely used to calculate the geometric phase shift acquired by a particle moving along a closed path---a phase shift that is determined only by the geometry of the path and is independent of the time spent \textit{en route}. Geometric phases \editout{also} provide an elegant description of the celebrated Aharonov-Bohm effect \cite{AharonovBohm1959}, where a magnetic flux in a confined region of space influences the eigenstates everywhere via the magnetic vector potential. In condensed-matter physics, an analogous Berry flux in momentum space is responsible for various anomalous velocities and Hall responses \cite{Xiao2010} and lies at the heart of many-body phenomena ranging from quantum Hall physics \cite{Thouless1982} to topological insulators \cite{Hasan2010}. The Berry flux density (Berry curvature) is indeed essential to the characterization of an energy band and determines its topological invariants. However, mapping out the geometric structure of an energy band \cite{Price2012,Dauphin2013,Abanin13} has remained a major unresolved challenge for experiments.

\begin{figure}[htb]
	\centering
		\includegraphics[width=84mm]{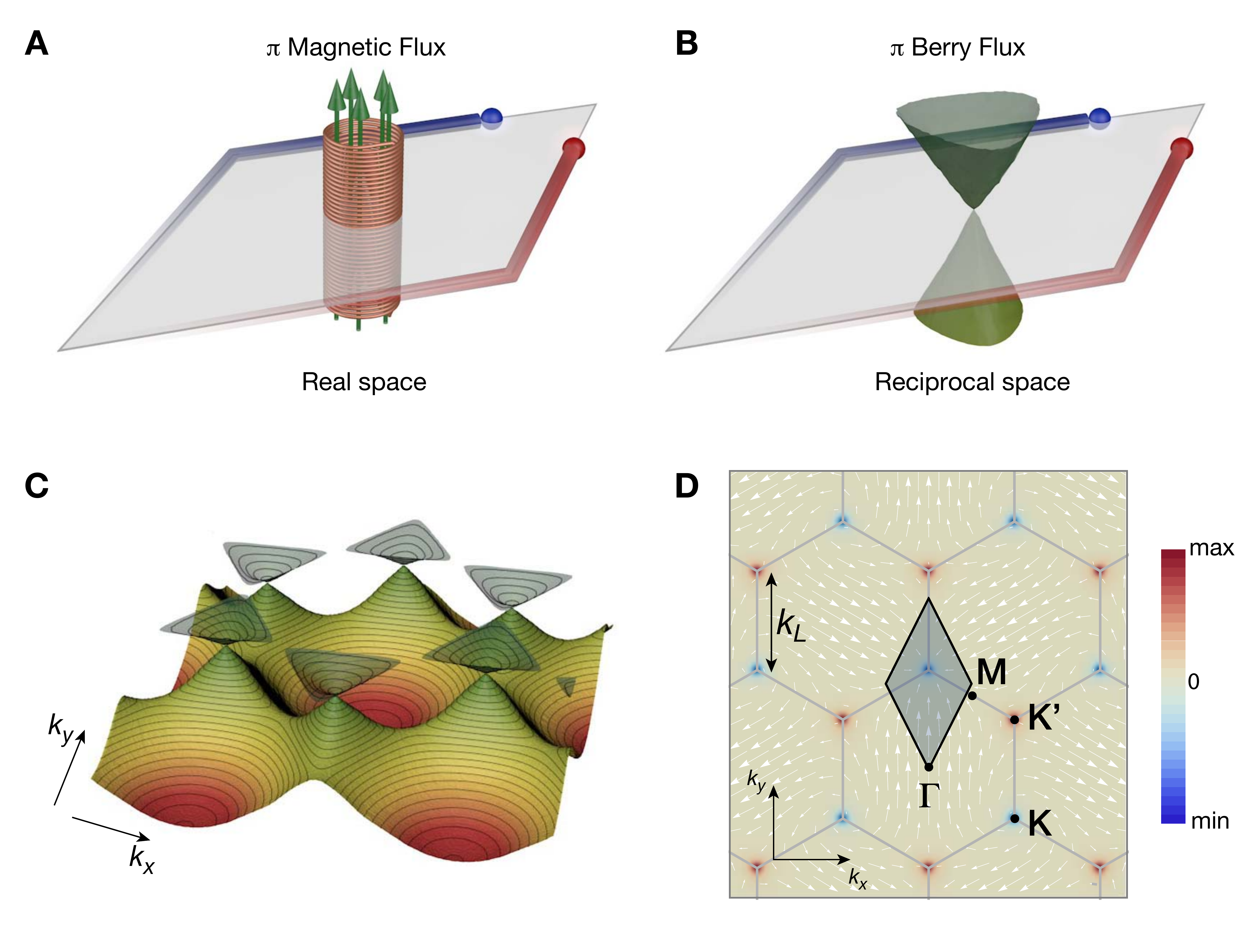}
    \caption{\textbf{Aharonov-Bohm analogy and geometric properties of the hexagonal lattice.} In the Aharonov-Bohm effect (\textbf{A}), electrons encircle a magnetic flux in real space, while in our interferometer (\textbf{B}), the particles encircle the $\pi$ Berry flux of a Dirac point in reciprocal space. In both cases, the flux through the interferometer loop gives rise to a measurable phase. \textbf{C}, Dispersion relation of the hexagonal lattice, showing the conical intersection between the first and second band at the Dirac points. \textbf{D}, Berry curvature of the first band calculated in the tight-binding regime with a gap of $\Delta$=0.5$J$ for visualization purposes. Dirac points are located at the corners ($\mathbf{K}$ and $\mathbf{K'}$ points) of the Brillouin zone (BZ) (gray hexagons). White arrows are a pseudo-spin representation of the Bloch states with orientation indicating the phase of the coupling between sublattices; lengths of the arrows indicate the energy gap in the two-band model. 
Also shown is a typical interferometer path (black diamond).
}
	\label{Fig:1}
\end{figure}

\begin{figure*}[tbh]
	\centering
		\includegraphics[width=160mm]{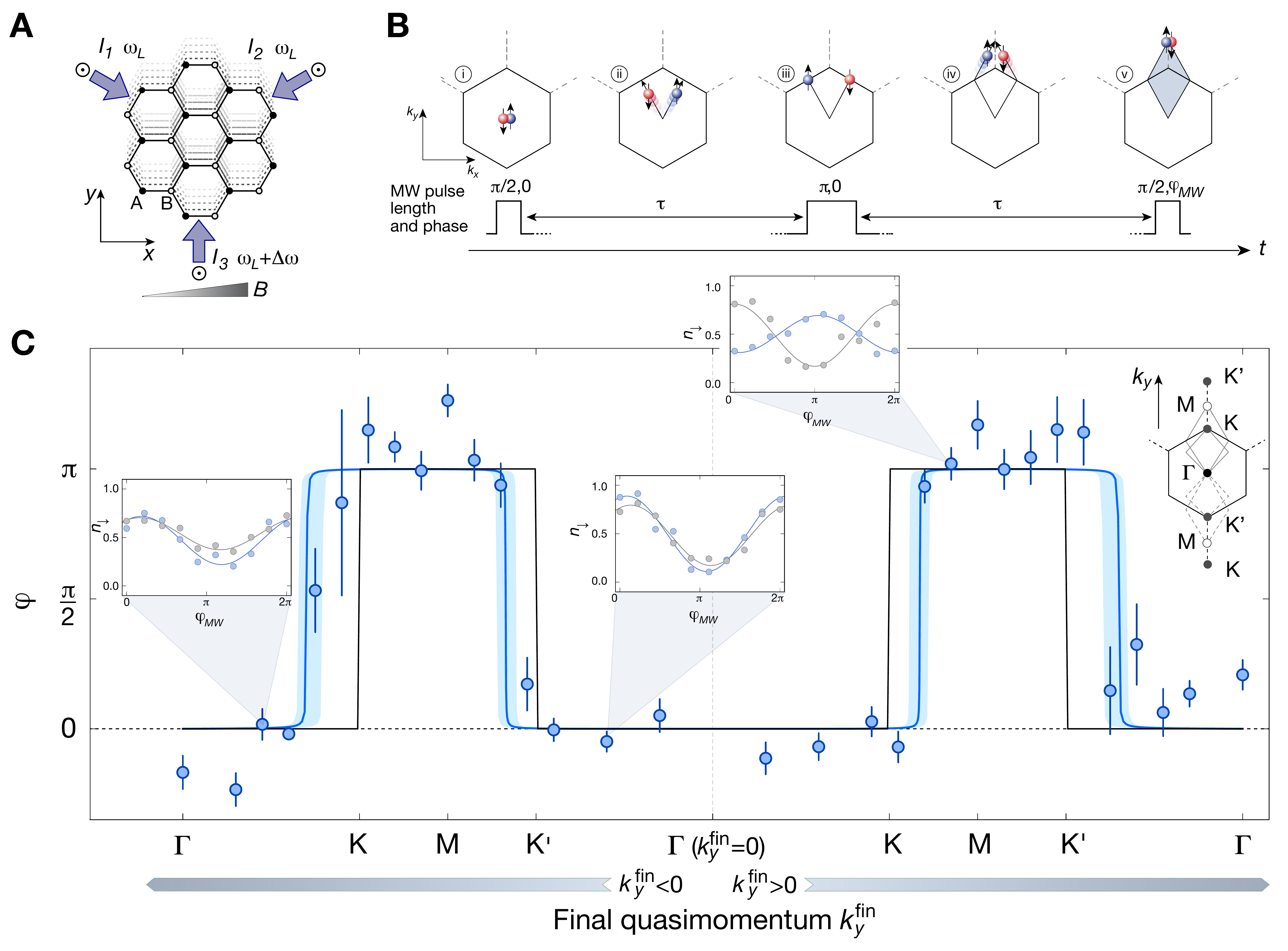}
    \caption{\textbf{Momentum-resolved detection of Berry flux at the Dirac points.} \textbf{A}, Sketch of the hexagonal lattice in real space with A (B) sites denoted by solid (open) circles. The lattice is realized by interfering three laser beams (blue arrows) of wavelength $\lambda_{\text{L}}$, intensity $I_\text{i}$, and frequency $\omega_{\text{L}}$, with linear out-of-plane polarizations. A linear frequency sweep of lattice beam three creates a uniform lattice acceleration along the $y$-direction. A magnetic field gradient $B'=9.0(1)\, $G/cm along the $x$-axis creates an additional spin-dependent force. \textbf{B}, Interferometer sequence. Hexagons indicate the first Brillouin zone and red (blue) spheres are atoms in the \smash{$|\!\downarrow\rangle$} (\smash{$|\!\uparrow\rangle$}) state. The duration of the interferometer sequence is \smash{$2\tau=1.6\,\text{ms}$} for all measurements. \textbf{C}, Summary of phase differences between measurement and reference loop for different final quasimomenta $k_y^{\text{fin}}$. Error bars denote fit uncertainties or standard deviations in case of averages. Lines are \textit{ab initio} theory using a full band structure calculation with: no momentum spread \smash{$\sigma_{\mathbf{k}}=0$} and perfectly localized Berry curvature $\ww=0$ (black); or $\sigma_{\mathbf{k}}$=0.21$k_L$ and $\ww\simeq 10^{-4}k_L$ (blue). The shaded area accounts for an experimental uncertainty of \smash{$\sigma_{\mathbf{k}}=0.14$--$0.28k_L$}. Insets show the fraction of atoms $n_\downarrow$ measured as a function of the phase $\varphi_{\text{MW}}$ for selected quasimomenta. Measurement loop data are shown in blue and reference loop data are shown in gray with corresponding sinusoidal fits.}
	\label{Fig:2}
\end{figure*}

Here, we demonstrate a versatile technique for measuring geometric phases in reciprocal space using spin-echo interferometry with ultracold atoms \cite{Abanin13,Atala13}. In contrast to typical solid state experiments, where all geometric effects are averaged over the Fermi sea, the use of a Bose-Einstein condensate (BEC) enables measurements with high momentum resolution. We exploit this resolution to directly detect the singular topological properties of an individual Dirac cone \cite{Neto09} in a graphene-type hexagonal optical lattice (see Fig.\,\ref{Fig:1}). Concentrated at the Dirac point is a $\pi$ Berry flux, which is analogous to a magnetic flux generated by an infinitely narrow solenoid \cite{Mikitik1999}. This localized flux gives rise to several striking properties of graphene, including the half-integer shift in the positions of quantum Hall plateaus \cite{Zhang2005,Novoselov2005}, the phase of Shubnikov-de Haas oscillations \cite{Zhang2005,Novoselov2005}, and the polarization dependence in photoemission spectra \cite{Liu11,Hwang11}. A similar $\pi$ flux also plays a crucial role in the nuclear dynamics of molecules featuring conical intersections of energy surfaces \cite{Wilczek89}. Our direct detection of the paradigmatic $\pi$ flux demonstrates the capability to reveal even singular Berry flux features that are not observable by alternative techniques based on transport measurements \cite{Price2012,Dauphin2013,Aidelsburger2014,Jotzu2014,Spielman2013} and thereby paves the way to full topological characterization of optical lattice systems \cite{Aidelsburger2011,Spielman2013,Tarruell2012,Struck13,Jotzu2014,Aidelsburger13,Miyake13,Goldman2013}.

The effect of Berry curvature in our interferometer is analogous to the Aharonov-Bohm effect, where an electron wavepacket is split into two parts that encircle a given area in real space (see Fig.\,\ref{Fig:1}A). Any magnetic flux through the enclosed area gives rise to a measurable phase difference between the two components. This remains true even if the magnetic field vanishes everywhere along the paths, and thus exerts no mechanical force on the electron. For a single Bloch band in the reciprocal space of a lattice system, an analog of the magnetic field is the Berry curvature $\Omega_n$ (see Eq.\ 1), which we probe by forming an interferometer on a closed path in reciprocal space (see Fig.\,\ref{Fig:1}B). The geometric phase acquired along the path can be calculated from the Berry connection $\vec{A}_n$, the analog of the magnetic vector potential. For a lattice system with Bloch waves $\psi_{ \k}^n( \rr)=e^{i \k\rr}u_{\k}^n(\rr)$ with quasimomentum $\k$ in the $n^{\text{th}}$ band and the cell-periodic part of the wave-function $u_{\k}^n(\rr)$, the Berry connection is given by \smash{$ \text{ \textbf{A}}_n( \k)=i \langle u_{\k}^n| \nabla_\k|u_{\k}^n\rangle$}. Accordingly, the phase along a closed loop in reciprocal space is
\begin{equation}
\label{eq:berryphase}
\varphi_{\text{Berry}} = \oint_C \text{ \textbf{A}}_n (\k)\, d \k=\int_S \Omega_{n}(\k)\, d^2k
\end{equation}
where $S$ is the area enclosed by the path $C=\partial_S$, and $\Omega_n=\nabla_{\k} \times \text{ \textbf{A}}_n( \k)$ the Berry curvature (color shading in Fig.\,\ref{Fig:1}D) \cite{Xiao2010}.
Although neither the magnetic vector potential nor the Berry connection is uniquely defined, the geometric phase acquired along a closed loop is gauge independent \cite{Berry1984}, and is therefore a measurable observable that encodes information on the geometrical properties of a Bloch band.

We implemented the graphene-like hexagonal optical lattice for ultracold $^{87}$Rb atoms by superimposing three linearly polarized blue-detuned running waves at 120(1)$^{\circ}$ angles, as depicted in Fig.\,\ref{Fig:2}A. The resulting dispersion relation includes two non-equivalent Dirac points with opposite Berry flux located at $ \KK$ and $\KKp$, which are repeated in every BZ (see Fig.\,\ref{Fig:1}D). The origin of the $\pi$ Berry flux lies in the bipartite structure of the hexagonal lattice \cite{Neto09}. As the unit cell contains two non-equivalent lattice sites A and B (Fig.\,\ref{Fig:2}A), the Bloch wave of the lowest band has the form of a spinor. Time-reversal and inversion symmetries constrain this spinor to the equatorial plane of its Bloch sphere, and each Dirac point acts as a magnetic monopole in reciprocal space \cite{Berry1984,Mikitik1999}, giving rise to the winding phase of the spinor in Fig.\,\ref{Fig:1}D. The spinor Bloch wave, just as a real spin-1/2 particle in a slowly rotating magnetic field, therefore acquires a geometric phase of $\pi$ along any trajectory enclosing a single Dirac point. The Berry curvature is thus confined to a perfectly localized $\pi$ Berry flux, $\Omega_n=\pm \pi\delta(\k-\KK^{(')})$ (see supporting online material, SOM), provided the aforementioned symmetries hold. Generically, the inversion symmetry may be broken by a slight ellipticity of the lattice beam polarizations, which introduces a small energy offset $\Delta$ between the A and B sites \cite{Hasan2010}. Such an offset opens a small gap at the Dirac points and spreads the Berry curvature over a finite range of quasimomenta, as shown in Fig.\,\ref{Fig:1}D. By probing for a spread in Berry curvature, we can thus place a bound on imperfections in the lattice, while simultaneously benchmarking the resolution of our interferometer.

\begin{figure}[htb]
	\centering
		\includegraphics[width=84mm]{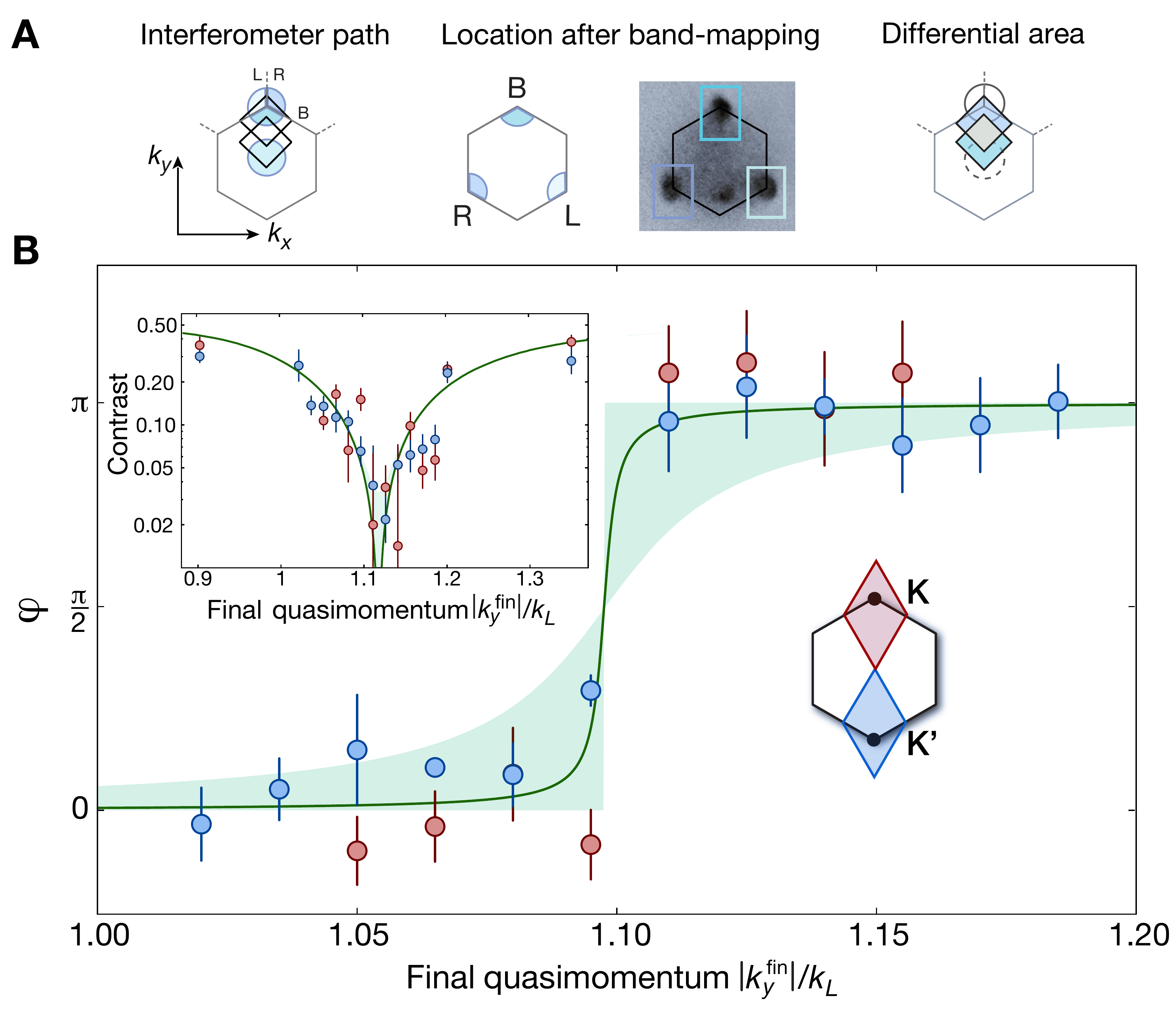}
    \caption{\textbf{Self-referenced interferometry at the Dirac point.} \textbf{A}, Examples of interferometer paths closing near the edge of the BZ for different initial quasimomenta: Due to the initial momentum spread, the cloud (circle with colored sectors, not to scale) is split by the edges of the BZ. Band-mapping spatially separates the three different parts of the cloud onto three corners of the first BZ (schematic and image). The measured differential Berry phase between the parts is the integral of $\Omega_n$ over non-common surfaces (blue-shaded areas). \textbf{B}, Phase differences between atoms that have crossed the band edge and the lagging (bottom) cloud versus final quasimomentum $k_y^{\text{fin}}$ for paths close to the $\KK$ ($ \KKp$) point in red (blue). The shaded region indicates a range of $\ww=0-12\times 10^{-4}k_L$ for the Berry curvature, while the line is calculated for \smash{$\ww\simeq 10^{-4}k_L$}, corresponding to an A-B offset of \smash{$\Delta=3\times 10^{-4}E_r$}. The inset shows the contrast $(n_\downarrow^{\text{max}}-n_\downarrow^{\text{min}})/({n_\downarrow^{\text{max}}+n_\downarrow^{\text{min}}})$ of the interference fringes of the full cloud. Theory line and shading is identical to the main graph. All calculations assume $\sigma_\k=0.15k_L$.
}
	\label{Fig:3}
\end{figure}

The interferometer sequence (see Fig.\,\ref{Fig:2}B) begins with the preparation of an almost pure $^{87}$Rb BEC in the state \smash{$|\!\uparrow\rangle=|F=2,m_{F}=1\rangle$} at quasimomentum $\k=0$ in a $V_0=1\,E_r$ deep lattice, where \smash{$E_r= \text{h}^2/(2m\lambda_L^2)\approx \text{h}\times 4 \,\text{kHz}$} is the recoil energy, and $\text{h}$ is Planck's constant. A resonant $\pi/2$-microwave pulse creates a coherent superposition of \smash{$|\!\uparrow\rangle$} and \smash{$|\!\downarrow\rangle=|F=1,m_F=1\rangle$} states (i). Next, a spin-dependent force from a magnetic field gradient and an orthogonal spin-independent force from lattice acceleration move the atoms adiabatically along spin-dependent paths in reciprocal space (ii) \cite{Dahan1996}. The two spin-components move symmetrically about a symmetry axis of the dispersion relation.
After an evolution time $\tau$, a microwave $\pi$-pulse swaps the states \smash{$|\!\downarrow\rangle$} and \smash{$|\!\uparrow\rangle$} (iii). Each atomic wavepacket now experiences an opposite magnetic force in the $x$-direction, such that both spin components arrive at the same quasimomentum $\k^{\text{fin}}$ after an additional evolution time of $\tau$ (iv). At this point, the state of the atoms is given by \smash{$|\psi^{\text{fin}}\rangle\propto|\! \uparrow, \k^{\text{fin}}\rangle+e^{i \varphi}|\!\downarrow, \k^{\text{fin}}\rangle$} with relative phase $\varphi$. A second $\pi$/2-microwave pulse with a variable phase $\varphi_{MW}$ closes the interferometer (v) and converts the phase information into spin population fractions \smash{$n_{\uparrow,\downarrow}\propto1\pm\cos(\varphi+\varphi_{MW})$}, which are measured by standard absorption imaging after a Stern-Gerlach pulse and time-of-flight.

The phase difference $\varphi$ at the end of the interferometer sequence consists of the geometric phase and any difference in dynamical phases between the two paths of the interferometer. Ideally, the dynamical contribution should vanish due to the symmetry of the paths and the use of the spin-echo sequence (see SOM). To ascertain that the measured phase is truly of geometric origin, we perform a reference measurement with a 'zero-area' interferometer, comprising a V-shaped path produced by reversing the lattice acceleration after the $\pi$-microwave pulse of Fig.\,\ref{Fig:2}B (iii).

As a key result of this work, we locate the Berry flux of the Dirac cone by performing a sequence of interferometric measurements in which we vary the region enclosed by the interferometer. This is achieved by varying the lattice acceleration at constant magnetic field gradient to control the final quasimomentum $k_y^{\text{fin}}$ ($k_x^{\text{fin}}=0$) of the diamond-shaped measurement loop. The resulting phase differences between measurement and reference loops are shown in Fig.\,\ref{Fig:2}C. When one Dirac point is enclosed in the measurement loop, we observe a phase difference of $\varphi \simeq \pi$. In contrast, we find the phase difference to vanish when enclosing zero or two Dirac points. We find very good agreement between our data and theoretical predictions (see Fig.\,\ref{Fig:2}C) accounting for the momentum spread of the BEC. This momentum spread affects the positions of the $\pi$ phase jumps but does not limit their sharpness. Indeed, the data are consistent with the step function expected for an inversion-symmetric lattice, where it is impossible to identify the sign of the Berry flux ($\pm\pi$). Small deviations of the phases from $0$ or $\pi$ can be due to an imperfect alignment of the magnetic field gradient, magnetic field fluctuations or an imperfect lattice geometry (see SOM).

To minimize systematic errors and improve our measurement precision, we performed self-referenced interferometry close to the Dirac points. Due to the finite initial momentum spread of the BEC, each atom moves along a slightly different measurement loop, as shown in Fig.\,\ref{Fig:3}A. As in Bloch oscillation experiments \cite{Dahan1996}, a standard band-mapping technique \cite{Greiner2001} projects those 'slices' of the cloud that have (left, right) or have not (bottom) crossed the edge of the BZ onto three different corners of the first BZ (see Fig.\,\ref{Fig:3}A), such that we can measure their acquired phases independently. This method effectively uses the edge of the first BZ to increase the quasimomentum resolution. By using the phase of the Ramsey signal from the lagging bottom cloud as a reference phase $\varphi_B$, we also reduce sensitivity to drifts in the experiment. The measured phase differences \smash{$\varphi=(\varphi_L+\varphi_R)/2-\varphi_B$}, where $\varphi_L$ and $\varphi_R$ refer to the phases of the left and right clouds in the BZ, respectively, show a sudden jump from 0 to $\pi$ as the atoms cross the edge of the band (see Fig.\,\ref{Fig:3}B). The position of the phase jump is in excellent agreement with a numerical calculation including an initial momentum spread of \smash{$\sigma_{\mathbf{k}}= 0.15(1)k_L$}, consistent with an independent time-of-flight measurement. Remarkably, the phase jump occurs within a very small quasimomentum range of $<0.01\, k_L$, and an arctangent fit to the experimental data gives a phase difference of $\varphi=$0.95(10)$\pi$. Both results are compatible with a perfectly localized and quantized $\pi$ Berry flux. 

To constrain the possible spread in Berry curvature, we analyze not only the phase (Fig.\, \ref{Fig:3}B) but also the contrast of the interference fringes, plotted in the inset of Fig.\,\ref{Fig:3}B. The location of the Dirac cone manifests itself through a pronounced minimum in the interference contrast. The sharpness of the phase jump and the strong reduction of contrast down to our detection limit demonstrate that the interferometric protocol can map the Berry curvature with extraordinarily high resolution. In the case analyzed here, the contrast measurements provide an upper bound for the spread of the Berry curvature around the Dirac cone of $\ww<6\times10^{-4} k_L $ (HWHM), corresponding to a maximal A-B site offset of $\Delta<\text{h}\times$12\,Hz and a ratio of energy gap at the Dirac cone to bandwidth of $<3\times10^{-3}$. Remarkably, the steepness of the phase jump in Fig.\,\ref{Fig:3}B suggests an even stronger localization of the Berry curvature on the order of $\ww\simeq10^{-4} k_L $ ($\Delta\simeq\text{h}\times$3\,Hz).

\begin{figure}[htb]
	\centering
		\includegraphics[width=84mm]{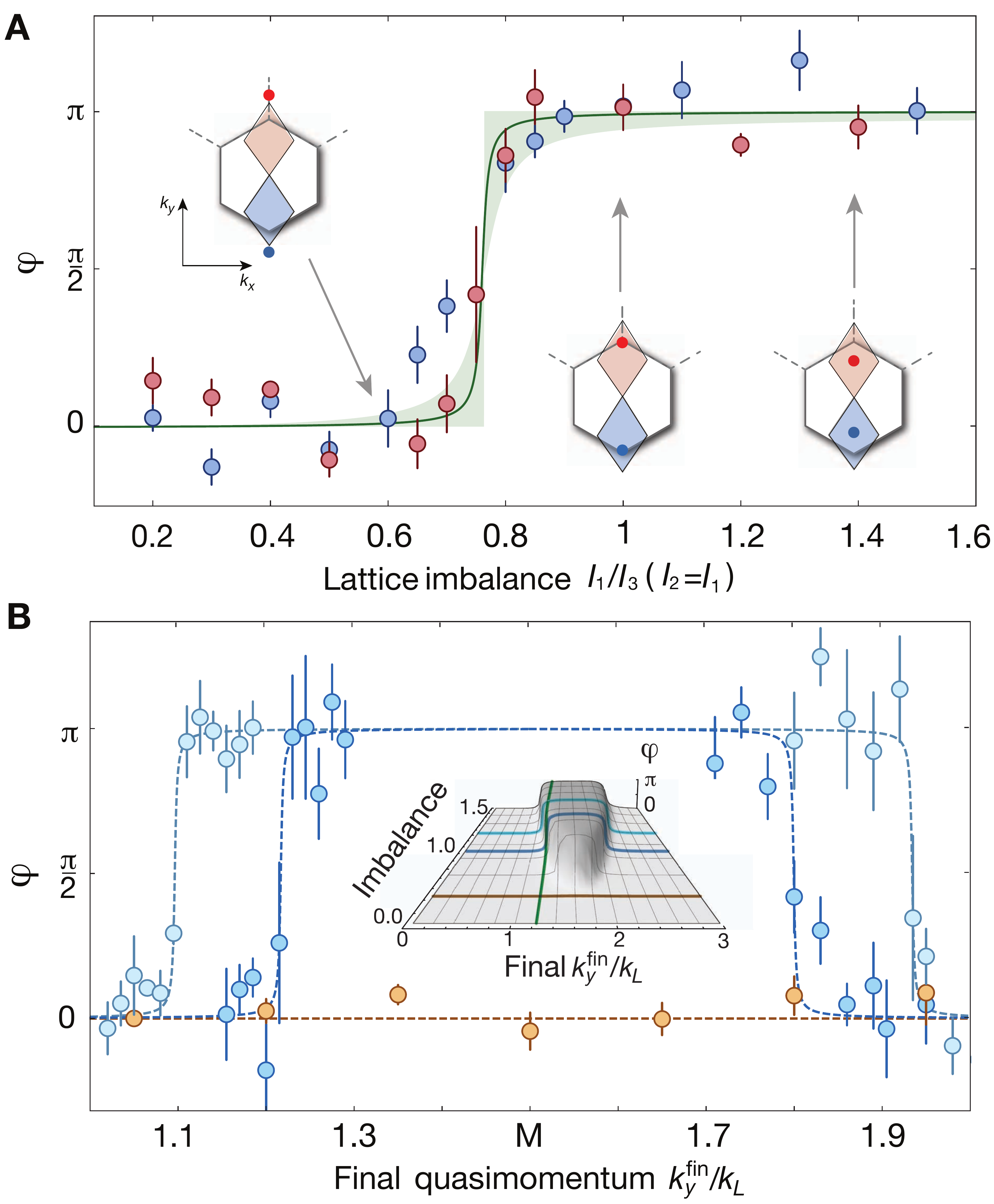}
    \caption{\textbf{Mapping the movement of Berry flux under distortion of the lattice.} \textbf{A}, Phase difference between reference and measurement loop versus lattice imbalance around $\KK$ (red) and $\KKp$ (blue) for a fixed final quasimomentum $k_y^{\text{fin}}=\pm 1.2k_{L}$. Red and blue dots in the insets give the location of Dirac points for the indicated imbalances. Theory curve is calculated for lattice depth $V_0=1E_r$, momentum spread $\sigma_\k$=0.15$k_L$, and  \smash{$\ww\simeq 10^{-4}k_L$}. Shaded area corresponds to  \smash{$\ww=0-12\times 10^{-4}k_L$}. \textbf{B}, Self-referenced phase near $\KK$ and $\KKp$ for an imbalance $I_{1,2}/I_3=1.0$ and $I_{1,2}/I_3=0.7$ in light and dark blue, highlighting the shift in the location of Berry flux. Phases are measured as in Fig.\,\ref{Fig:3}. Orange data are phase differences between the measurement and reference loops for an imbalance of $I_{1,2}/I_3=0.2$, where no phase shift is observed. Curves are guides to the eye. The inset shows the calculated Berry phase for loops with various final $k_y^{\text{fin}}$ and lattice imbalances using the same $\sigma_\k$ and $\ww$ as above. Colored lines indicate parameters explored in the measurements.
 }
	\label{Fig:4}
\end{figure}

To verify the method's sensitivity to changes in Berry flux, we performed interferometry in a modified lattice potential. Changing the power of two lattice beams ($I_1,I_2$) relative to the third ($I_3$) deforms the lattice structure, while preserving time-reversal and inversion symmetry. With decreasing $I_{1,2}/I_3<1$, the Dirac points and the associated fluxes move toward each other along the symmetry axis of the interferometer loop \cite{Neto09} (insets of Fig.\,\ref{Fig:4}A). Nonetheless, the Berry flux singularities remain protected by symmetry until the Dirac points merge and annihilate \cite{Tarruell2012}. By using a fixed measurement loop that encloses one Dirac point in the intensity-balanced case, we can measure the change of the geometric phase as we imbalance the lattice beam intensities. The measured Berry phases drop from $\pi$ to 0 as the Dirac point moves out of the loop, in very good agreement with \textit{ab initio} calculations (see Fig.\,\ref{Fig:4}A).
To precisely map the location of the Berry flux in the imbalanced lattice, we again use the self-referenced interferometry of Fig.\,\ref{Fig:3}. As shown in Fig.\,\ref{Fig:4}B, imbalancing the lattice by decreasing $I_{1,2}/I_3$ narrows the range of final quasimomenta for which the interferometer encloses a single $\pi$ flux, thereby shifting both the upward and downward phase jumps towards the $\vec{M}$ point. The position of the phase jump at $\kyfin=1.2 k_L$ for $I_{1,2}/I_3=0.7$ is in very good agreement with theory, while deviations of $\simeq $10\% from the calculated value in the positions of the phase jumps at higher quasimomenta can likely be attributed to a combination of geometric imperfections and the effect of the dynamical instability of the Gross-Pitaevskii equation \cite{Fallani2004}. For a stronger imbalance ($I_{1,2}/I_3$=0.2), the two Dirac points have annihilated, and hence no phase jump is observed for any loop size.

To conclude, we have demonstrated the first momentum-resolved measurement of Berry curvature, using atom interferometry. By employing a Bose-Einstein condensate, we constrain the spread of Berry curvature around a Dirac point in a hexagonal optical lattice to be smaller than $\ww<10^{-3} k_L$, thereby highlighting both the singular topological nature of a conical intersection and the resolution achievable with this method. Our Aharonov-Bohm-type interferometer allows one to fully resolve the geometric structure of a single Bloch band by combining local measurements of the Berry phase along small paths, thereby enabling the full reconstruction of topological invariants such as Chern numbers. The method can readily be applied to a variety of optical lattices and other physical settings such as polariton condensates \cite{Carusotto2013}. Multiband extensions of this work can enable measurements of Wilson loops and off-diagonal (non-Abelian) Berry connections and thus provide a framework for the full determination of the geometric tensor of Bloch bands in periodic structures \cite{Grusdt2014}. Controlled application of non-Abelian Berry phases would furthermore constitute a key step towards holonomic quantum computation \cite{Zanardi99}. Even within a single topologically trivial band, the possibility of preparing a BEC or Fermi-sea at finite quasimomentum should enable the observation of transient Hall responses due to local Berry curvature and, combined with the possibility of performing quantum quenches and the control of interactions, is expected to lead to novel many-body phenomena \cite{Parameswaran2013}. Finally, the highly non-linear phase jump we have observed at the Dirac point may find application in precision force sensing \cite{Tuchman2009}.

%

\section*{Acknowledgements}
We acknowledge technical assistance by M. Boll, H. L\"{u}schen and J. Bernardoff during the setup of the experiment and would like to thank E. Demler, D. Abanin and X.-L. Qi for helpful discussions. We acknowledge financial support by the Deutsche Forschungsgemeinschaft (FOR801), the European Commision (UQUAM), the U.S. Defense Advanced Research Projects Agency Optical Lattice Emulator program, and Nanosystems Initiative Munich.
\noindent

\vspace{0.5cm}

\clearpage
\renewcommand{\thefigure}{S\arabic{figure}}
\renewcommand{\theequation}{S.\arabic{equation}} 
\renewcommand{\citenumfont}[1]{S#1}
\renewcommand{\bibnumfmt}[1]{[S#1]}
\renewcommand{\thesection}{S\Roman{section}}
\renewcommand{\thesubsubsection}{\roman{subsubsection}}

\setcounter{equation}{0}
\setcounter{figure}{0}

\section*{Supporting material}


This appendix provides the theoretical background for our interferometric characterization of band topology in a hexagonal optical lattice and additional experimental details. In Section \ref{sec:bloch_interferometry}, we present the theory of the Aharonov-Bohm-type interferometer in momentum space. Section \ref{sec:tightbinding} reviews the origin of the Berry flux in a honeycomb lattice, focusing on a tightbinding model that captures the essential physics. We proceed in Sec.\ \ref{sec:actual_lattice} to a complete description of the honeycomb lattice as realized in our experiment. Finally, in Sec.\ \ref{sec:momentum_spread} we account for effects of the atomic momentum distribution to verify our detailed quantitative understanding of the experimental results. In Sec.\ \ref{sec:2}, we provide additional experimental details.

\section{Theoretical Background}
\subsection{Aharonov-Bohm Interferometry in 2D Bloch Bands}\label{sec:bloch_interferometry}
To form an interferometer in reciprocal space, we combine a magnetic field of magnitude $B=B_0+\vec{r}\cdot\nabla B$ with  an orthogonal acceleration \smash{$\aa\perp \nabla B$} of the lattice. The resulting time-dependent Hamiltonian for an atom of magnetic moment $\mu$ and mass $m$ is
\begin{equation}
H(t) = \frac{\p^2}{2m}+V\left[\rr-\Rr\left(t\right)\right] - \mu\, \rr\cdot\nabla B - \mu B_0,
\end{equation}
where $V(\rr)$ describes the lattice potential at $t=0$ and $\Rr(t)=\aa t^2/2$. The dynamics of this Hamiltonian is most conveniently analyzed in a frame co-moving with the lattice, which we enter via a unitary transformation $U(t) = e^{-i\rr\cdot m\aa t}e^{i \p\cdot\Rr(t)}$, with $\hbar=1$. The time-dependent Schr\"{o}dinger equation $i\dot{\Psi}=H(t)\Psi$ can then equivalently be expressed as $i\dot{\tpsi}=\tilde{H}\tpsi$, where $\tpsi = U\Psi$ and
\begin{align}\label{eq:Hlattice}
\tilde{H} = UHU^\dagger + i\dot{U}U^\dagger &= \frac{\vec{p}^2}{2m}+V(\rr) - \fs \cdot\rr + \varepsilon_\mu(t).
\end{align}
Here, $\p^2/(2m)+V(\rr)\equiv H_0$ is the bare lattice Hamiltonian, $\fs=\mu\nabla B-m\aa$ includes both the magnetic force and the fictitious force experienced by the atoms in the non-inertial lattice frame, and $\varepsilon_\mu(t)=-\mu[\Rr(t)\cdot\nabla B+B_0]$ describes the Zeeman energy. The $\vec{R}$-dependent Zeeman contribution is ideally kept zero by setting the acceleration to be orthogonal to the magnetic field gradient. The effect of the energy $\mu B_0$ is removed by our spin-echo sequence, provided that the magnetic field is constant over the duration of the experiment. We nevertheless retain $\varepsilon_\mu$ in our analysis to remain aware of potential sources of experimental error. We omit in Eq.\ \ref{eq:Hlattice} a kinetic energy offset $\frac{1}{2}m|\aa t|^2$ that is common to both spin states.

The effect of the force $\fs$ is to induce a translation $\k\rightarrow \k + \fs t$ in reciprocal space. To verify this, and to calculate the phase acquired in the process, we substitute into the time-dependent Schr\"{o}dinger equation the ansatz
\begin{equation}
\tpsi(t)=e^{i\eta(t)}\psi^n_{\k_0+\fs t},
\end{equation}
where $\psi^n_{\k}(\rr)=e^{i\k\cdot\rr}u^n_{\k}(\rr)$ are Bloch wavefunctions satisfying $H_0\psi^n_{\k}(\rr)=E_n(\k)\psi^n_{\k}(\rr)$ for the $n^{th}$ band. We assume that the force is sufficiently weak to restrict the dynamics to a single band of index $n=1$, a condition satisfied in our experiment (see Sec.\ \ref{sec:2}). After a time $\tau$, the wave function acquires a phase $\eta = \phidyn + \phig$ that generically can include both a dynamical contribution
\begin{equation}\label{eq:phidyn}
\phidyn = \int_0^\tau\left[E_1\left(\k+\fs t\right) + \varepsilon_\mu\left(t\right)\right]\,dt
\end{equation}
and the geometric contribution that is our chief interest:
\begin{align}
\phig &= i\int_0^\tau\, \bra{u^1_{\k+\fs t}}\delk\ket{u^1_{\k+\fs t}}\cdot \fs\,dt \nonumber \\ &= i\int_C^{} \bra{u^1_{\k}}\delk\ket{u^1_{\k}}\cdot d\k.
\end{align}
The last equality emphasizes that, in contrast to the dynamical phase, $\phig$ depends \textit{only} on the path $C$ in reciprocal space and not on the time required to traverse it. Note that for an open path $C$, the geometric phase $\phig$ is gauge-dependent, as we are free to redefine the functions $u^n_{\k}$ by an arbitrary $\k$-dependent phase factor. Yet for any closed loop, such as is formed by our full spin-echo sequence (Fig.\ 2), $\phig$ is an observable, gauge-invariant quantity measuring the enclosed Berry flux \cite{Xiao:2010}.

To design an interferometer that measures only the Berry phase $\phig$ without dynamical phase contributions, we exploit the symmetry of the lattice under a reflection $\uvec{x}\rightarrow-\uvec{x}$. We choose the magnetic field gradient to lie along $\uvec{x}$ and the acceleration correspondingly along $\uvec{y}$. This ensures that two spin states of opposite magnetic moment sample the same dispersion relation at each point in time:
\begin{equation}\label{eq:dispersion_symmetry}
E_1\left(\k+\f{\abs{\mu}} t\right)=E_1\left(\k+\f{-\abs{\mu}} t\right).
\end{equation}
Ideally, the dynamical phase is thus common to both interferometer arms and has no influence on the measurement.

In practice, imperfections in alignment of the magnetic field gradient relative to the lattice, or errors in the relative angles or intensities of the lattice beams, can introduce small dynamical phases that contribute to our experimental uncertainty. For example, for the data in Fig.\  ~2  in the main text, the Zeeman term $\varepsilon_\mu(t)$ in the dynamical phase $\phidyn$ coming from an imperfect alignment of the magnetic field gradient increases linearly with the lattice acceleration, and consequently the final quasimomentum $\kyfin$, and is likely a dominant source of systematic error at large $\abs{\kyfin}$. The experimental tolerances on the alignment of the magnetic field gradient are discussed further in Section \ref{sec:2}.

\subsection{Berry Flux in a Hexagonal Lattice: Tight-Binding Model}\label{sec:tightbinding}

\begin{figure}[htb]
	\centering
		\includegraphics[width=84mm]{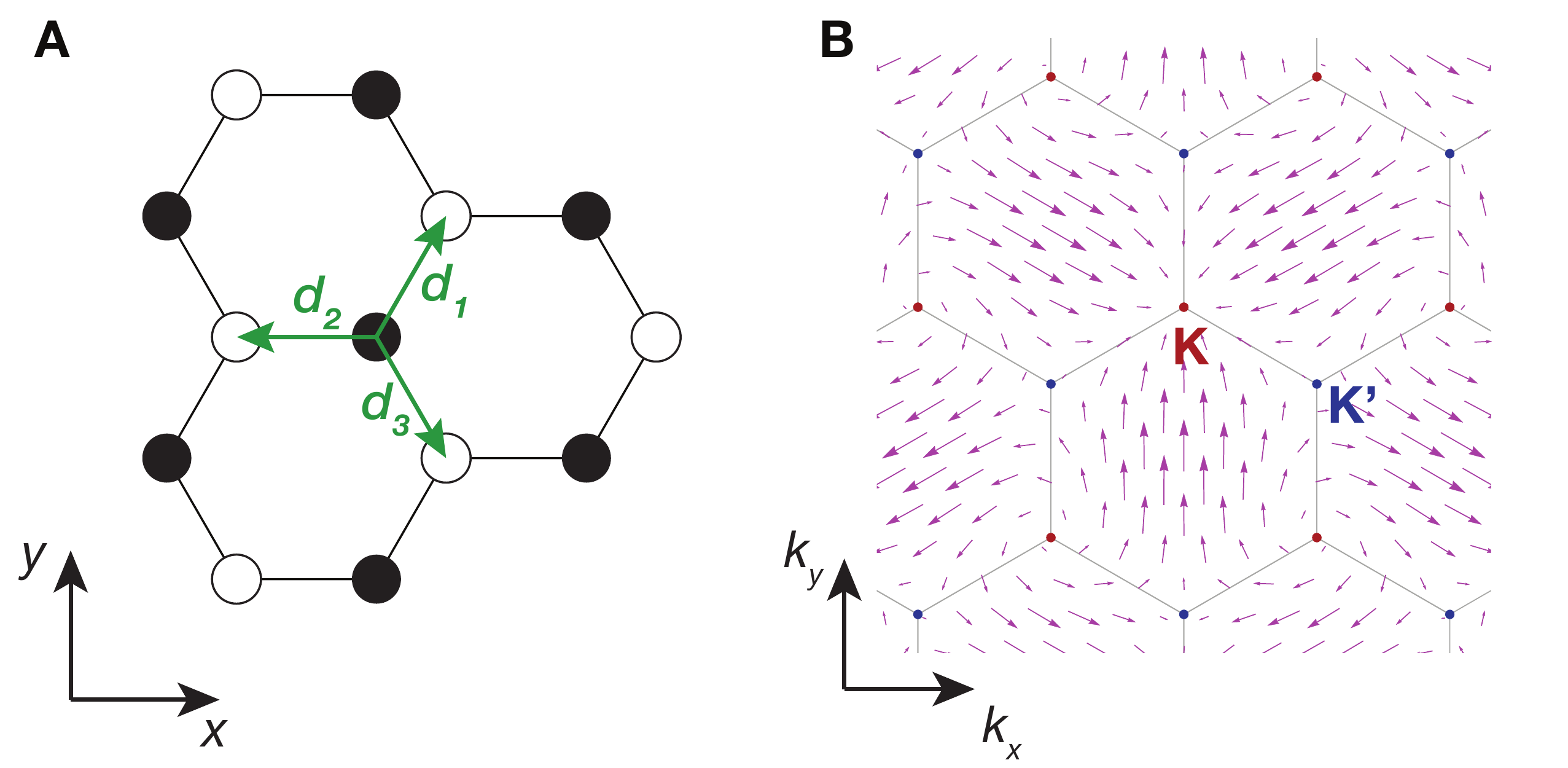}
    \caption{\textbf{Tight-binding model of the honeycomb lattice.} \textbf{A}, Lattice in real space, composed of sub-lattices $A$ (solid circles) and $B$ (open circles), with nearest-neighbor hopping vectors $\vec{d}_i$. \textbf{B}, Spinor eigenstates in reciprocal space, illustrated for the case of degenerate sublattices $\Delta=0$. Directions of the purple arrows indicate the orientation of $\vec{S}(\k)$ in the $x,y$-plane. Lengths of the arrows indicate the gap $E_+(\k) - E_-(\k)$ between eigenenergies $E_\pm$ of $\Htb$. The winding of $\vec{S}$ about the $\uvec{z}$ axis in the vicinity of each Dirac point $\KK, \KKp$ (red, blue) produces a sign change in the wave function of a particle that adiabatically encircles it \cite{Aharonov67}, corresponding to a Berry phase of $\pi$.}
\label{fig:s1}
\end{figure}

The origin of the Berry fluxes in the honeycomb lattice can readily be understood in the tight-binding limit, where the lattice may be decomposed into two triangular sublattices that are coupled by nearest-neighbor hopping (Fig.\ \ref{fig:s1}A). With the ground-state Bloch wavefunctions of the two sublattices as basis states, the two lowest bands of the honeycomb lattice are described by the Hamiltonian \cite{Semenoff84}
\begin{equation}
\label{eq:Htb}
\Htb(\k)=\begin{pmatrix}
\Delta/2 & -t_{\k} \\
- t_{\k}^{*} & -\Delta/2
 \end{pmatrix},
\end{equation}
where $\Delta$ is an energy offset between the sublattices and
\begin{equation}
\label{tkterm}
t_{\k}=Je^{i\k\cdot\dd_1}+Je^{i\k\cdot\dd_2}+Je^{i\k\cdot\dd_3},
\end{equation}
with $\mathbf{d}_i$ being  the nearest-neighbor lattice vectors and $J$ the hopping amplitude. The eigenstates of $\Htb$ are spinors $u_{\k}^\pm$, which may be visualized in terms of  the expectation value \smash{$\vec{S}(\k)=\pm\bra{\upm}\boldsymbol{\sigma}\ket{\upm}$} of the Pauli operator $\boldsymbol{\sigma}$ in the ground state. The momentum-dependent orientation of the pseudospin $\vec{S}(\k)$ is illustrated in Fig.\ \ref{fig:s1}B. For $\Delta=0$, the pseudo-spin $\vec{S}$ undergoes a full rotation in the $S_z=0$ plane in an infinitesimal loop around a Dirac point. This winding gives rise to the $\pi$ Berry flux at the Dirac point, in precise analogy to the $\pi$ phase acquired by a spin-1/2 particle as its alignment adiabatically follows a magnetic field through a single rotation in the $xy$-plane \cite{Aharonov67}.

More generally, for an arbitrary two-band system, the Berry curvature of the $n^\text{th}$ band may be calculated as
\begin{equation}
\Omega_n(\vec{k}) = i\delk\times\bra{u^n_{\k}}\delk\ket{u^n_{\k}} = \frac{\vec{S}}{2}\cdot\left(\frac{\partial\vec{S}}{\partial k_x}\times\frac{\partial\vec{S}}{\partial k_y}\right).
\end{equation}
For the honeycomb lattice with nearly degenerate sublattices ($\Delta/J \ll 1$), as in the case of our experiment, $\Omega$ is well approximated in the vicinity of each Dirac point $\KK_+\equiv\KK$ or $\KK_-\equiv\KKp$ by
\begin{equation}\label{eq:tb_bcurv}
\Omega_n(\vec{k})\approx \pm\frac{1}{2\gamma^2}\left(1+\abs{\frac{\k-\KK_\pm}{\gamma}}^2\right)^{-3/2},
\end{equation}
where $\gamma=\frac{1}{3 d}\frac{\Delta}{J}$ parametrizes the distribution of Berry curvature, and $d=\abs{\dd_i}$. We quantify the spread in Berry curvature in terms of the half-width at half maximum $\ww$ of the distribution $\Omega_n(\k)$. In the limit of perfect sublattice degeneracy ($\Delta=0$), Eq.\ \ref{eq:tb_bcurv} reduces to the singular form $\Omega_n(\vec{k})=\pm\pi\delta(\k-\KK_\pm)$. This singularity is imposed by the symmetries of the lattice under time reversal $\mathcal{T}$ and inversion $\mathcal{I}$, which preclude any loop in reciprocal space from enclosing a Berry flux with a well-defined sign \cite{Mikitik:1999}. 

\begin{figure*}[htb]
	\centering
		\includegraphics[width=168mm]{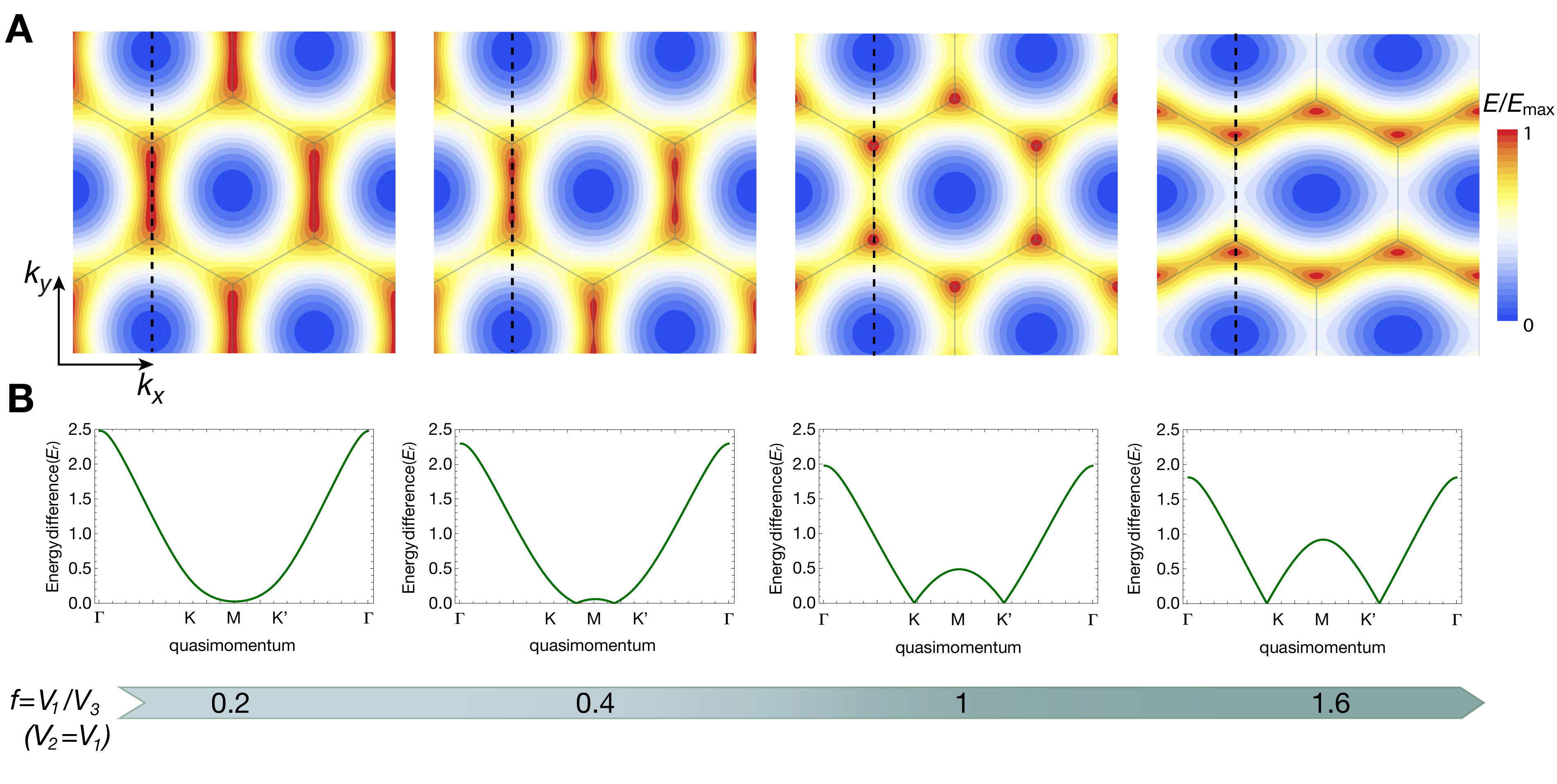}
    \caption{\textbf{Energy spectrum and movement of the Dirac points.} \textbf{A}, 2D plots of the lowest energy band from \emph{ab initio} calculations as a function of intensity imbalance $f=V_{1,2}/V_3$, where $V_3=1E_r$. The color scale on each plot is normalized to the band width $E_\mathrm{max}$. The dashed lines indicate the merging direction which is also the symmetry axis of the interferometer. \textbf{B}, Energy splitting between the two lowest bands for a cut along the merging direction ($k_x^{\text{fin}}$=0).
}
\label{fig:s2}
\end{figure*}

\subsection{Optical Hexagonal Lattice: Full Description}\label{sec:actual_lattice}

As our experiments are performed outside the tightbinding regime, we perform all theoretical modeling of the system by \textit{ab initio} band-structure calculations incorporating the full lattice potential. The experimental setup described in the main text produces a lattice of the form
\begin{align}
\label{eq:V}
V(x,y)=&\abs{\sum^3_{i=1}\sqrt{V_i}e^{-i\k_i\cdot\mathbf{r}}}^2 \nonumber \\ 
=&V_1+V_2+V_3+2\sqrt{V_1V_2}\cos(\sqrt{3}k_Lx)\nonumber \\
&+2\sqrt{V_1V_3}\cos\left(k_L\left(\frac{\sqrt{3}}{2}x-\frac{3}{2}y\right)\right)  \nonumber \\ &+2\sqrt{V_3V_2}\cos\left(k_L\left(\frac{\sqrt{3}}{2}x+\frac{3}{2}y\right)\right),
\end{align}
where $\k_i$ are the wave-vectors of the three lattice beams with wavenumber \smash{$k_{L}=\abs{k_i}$} and $V_{i}$ is the ac Stark shift produced by the $i^\mathrm{th}$ beam alone. Note that $V_i\propto I_i$, where $I_i$ are the intensities of the beams. All of our experiments are conducted with $V_1=V_2$, so that the lattice is symmetric under reflection $x\rightarrow -x$ about the symmery axis of the interferometer. This condition ensures, together with the orthogonality between lattice acceleration and magnetic field gradient,  that the interferometer measures only a geometrical phase while dynamical phases cancel out  (see Sec.\ \ref{sec:bloch_interferometry}).

\subsubsection{Imbalanced Lattice}
To vary the location of the Berry flux in the reciprocal lattice, we change the intensities of two lattice beams relative to the third, setting $V_{1,2}=fV_3$. The lower the imbalance factor $f$, the smaller the separation between the Dirac points along the $\uvec{y}$ direction becomes, as illustrated in Fig.\ \ref{fig:s2}. At a critically low imbalance factor $f_c$, the Dirac points merge and the corresponding Berry fluxes annihilate, leaving a gapped spectrum with no topological features for $f<f_c$ \cite{Tarruell:2012}. For the lattice depth of $1E_r$ employed in our experiments and considered in Fig.\ \ref{fig:s2}, $f_c=0.2$.

\subsubsection{Calculation of Berry curvature}
For the ideal honeycomb lattice defined in Eq.\ \ref{eq:V}, time-reversal and inversion symmetries dictate that the Berry curvature has to be localized in delta-function singularities. In practice, however, the Berry curvature may be spread out by experimental imperfections that break the inversion symmetry, such as ellipticity in the lattice beam polarizations \cite{Baur14}. To allow for a finite Berry curvature in our model, we add to the potential of Eq.\ \ref{eq:V} a term
\begin{equation}\label{eq:Vab}
V_{AB} = \frac{\Delta}{\sqrt{3}}\sin(\sqrt{3}k_L x)
\end{equation}
that introduces a small energy offset $\Delta$ between the $A$ and $B$ sublattices. 
By numerically diagonalizing the full Hamiltonian including this term, we calculate the Berry curvature from the eigenstates $\ket{u^n_{\k}}$ on a discrete grid in reciprocal space \cite{Fukui05}. Refining the grid via a local adaptive algorithm enables an efficient and precise calculation even for the highly localized Berry curvature in our system. To quantify the localization in Berry curvature and estimate the associated sublattice offset $\Delta$, we fit the numerical calculation with the model of Eq.\ \ref{eq:tb_bcurv}.

\subsection{Effects of Atomic Quasimomentum Distribution}\label{sec:momentum_spread}

To accurately relate the measured interferometer phases to the location of the Berry flux in reciprocal space, we must account for the quasimomentum spread of the atom cloud. In an interferometer that encloses a region $S$ for atoms initially at $\k=0$, an atom that instead has an initial quasimomentum $\vec{k}=\Q$ acquires a Berry phase
\begin{equation}
\Phi(\Q) = \int_S \Omega(\vec{k}+\Q)\,d^2\vec{k}.
\end{equation}
In our most straightforward analysis, we measure spin-state population fractions ($\nup$, $\ndown$) averaged over the entire cloud to obtain a Ramsey fringe
\begin{align}\label{eq:AvgFringe}
\nup-\ndown &= \int\cos\left[\phimw+\Phi(\Q)\right]n(Q)\,d^2\Q \nonumber\\ &= \mathscr{C} \cos(\phimw+\dphi),
\end{align}
where $n(\Q)$ represents the normalized initial quasimomentum distribution, with $\int n(Q)\,d^2\Q=1 $. The phase $\dphi=\arg(z)$ and the observed contrast $ \mathscr{C}\le\abs{z}$ are given by
\begin{equation}\label{eq:AvgZ}
z = \int n(\Q)\exp(i\Phi_\Q)\,d^2\Q.
\end{equation}
The actual interference contrast in the experiment is imperfect due to inhomogeneous broadening of the microwave transition and the heating associated with dynamical instability. In modeling the data, we therefore globally rescale the predicted contrast $ \mathscr{C}$ according to the maximum observed contrast for loops ending close to the Dirac point ($\Delta k\simeq 0.1-0.2\,k_L$), i.e., $\mathscr{C}\!=\!\mathscr{C}_\text{max} \times\abs{z}$ . We find excellent agreement between this simple model and our data.
Comparisons of experimental results with the above model are shown in Fig.\ 2-4 of the main text. In calculating each of the model curves in the graphs, we assume $n(\Q)$ to be Gaussian with standard deviation $\sigma_\k$. In Fig.\ 2, to account for heating during the sequence, the blue shaded area shows the predicted phase $\dphi$ vs.\ final quasimomentum for a range of values $0.14k_L\le\sigma_{\k}\le0.28k_L$. The minimum value of $\sigma_\k$ corresponds to the independently measured momentum spread of the cloud before the start of the interferometer sequence of $\sigma_\k=0.14(1)k_L$ (see the experimental section Sec.\ \ref{sec:atomprep} for details). Data in Fig.\ 3 and 4A, close to the edge of the BZ, are best fit with with a $\sigma_\k=0.15k_L$, which indicates at most a modest heating during the motion of the atoms in reciprocal space as they approach the first Dirac point.

\subsubsection{Auxiliary analysis near Dirac point}\label{sec:edge_analysis}

The self-referenced interferometry presented in Fig.\ 3 employs an auxiliary analysis of the images obtained for final quasimomenta in the vicinity of the Dirac point $\KK$. Here, the edges of the three Brillouin zones that touch at $\KK$ ``slice'' the atomic cloud into three components ($L,R,B$, as labeled in Fig.\ 3) that are well spatially separated in band-mapped pictures after time-of-flight (TOF) expansion. We perform independent fits to each of the three corresponding interference fringes to determine the phase
\begin{equation}\label{eq:slicing}
\dphi = (\dphi_L+\dphi_R)/2-\dphi_B.
\end{equation}
Here, $L$ and $R$ label the atoms in slices that are first to pass the Dirac point and thereby acquire a phase shift relative to atoms in cloud $B$. In modeling the self-referenced interferometer, we apply Eqs. \ref{eq:AvgFringe}-\ref{eq:AvgZ} to calculate the phase of each interference fringe, substituting for $n(\Q)$ one of the three slices of the full quasi-momentum distribution. The curve calculated for $\sigma_{\k}=0.15k_L$ fits the data in both Fig.\ 3B and Fig.\ 4B very well.

The data in Fig.\ 4B are obtained from self-referenced interferometry with paths enclosing up to two Dirac points. In the vicinity of the second Dirac point we again apply Eq.\ \ref{eq:slicing}, with the label $B$ now referring to the contingent of atoms that lead the procession along $\vec{k}_y$ and are thus first to sample the Berry flux of \text{both} Dirac points. Theory lines in Fig.\ 4B are arctangent fits with the slope of the phase jump fixed by our best estimate of the HWHM of the Berry curvature, $\ww\simeq 10^{-4}k_L$, from the data of Fig.\ 3B. 

\section{Experimental methods} \label{sec:2}

This section of the supplements provides additional information on the relevant experimental parameters and the optimizations taken to reduce potential sources of errors in the evaluation of the Berry phase.

\subsubsection{Preparation scheme }\label{sec:atomprep}

$^{87}$Rb atoms are cooled to quantum degeneracy by evaporative cooling initiated in a plugged quadrupole trap and completed in a crossed-beam dipole trap. The experimental sequence begins with an almost pure BEC of typically $4\cdot 10^4$ atoms in the internal state \smash{$|F=1,m_{F}=0\rangle$}. The magnetic field gradient is turned on 2s before the interferometer sequence starts to allow the current to stabilize. The atoms are then adiabatically loaded into a hexagonal lattice of chosen depth and configuration in 100\,ms. A 15$\mu$s microwave $\pi$-pulse transfers the atoms in $|1,0\rangle$ to $|2,1\rangle$ to start the spin-echo sequence.

Directly after loading into the lattice, the momentum spread measured via time of flight expansion is $\sigma_{\k}=0.14(1)k_L$. TOF images after the interferometry sequence show evidence of modest heating over the course of the motion, attributable to dynamical instabilities arising in regions of reciprocal space where the atoms acquire a negative effective mass  \cite{Fallani:2004}. 

\subsubsection{Lattice calibration and trap frequencies}

The lattice depth is calibrated via St\"{u}ckelberg interferometry \cite{Zenesini2010}. By measuring the energy difference between the first and second band at different locations in the Brillouin zone (BZ), we estimate a lattice depth of 1.0(1)$E_r$. The trap frequencies of the combined blue-detuned lattice and dipole potential are $\omega_{x,y}/2\pi$=26.5(7)~Hz and $\omega_{z}/2\pi$=183(2)~Hz. They are obtained by measuring the oscillation frequency of the center-of-mass motion of the BEC after a perturbation of the trapping potential. Due to the modest atom number and the rather small trap frequencies, the system is sufficiently dilute such that interaction effects can be neglected to first order.

\subsubsection{Detection} 

We perform a band-mapping sequence by linearly ramping down the lattice in 410 $\mu $s. During the 10 ms TOF, a Stern-Gerlach pulse of 9.5 ms is applied to separate the \smash{$|\!\uparrow\rangle$} and \smash{$|\!\downarrow\rangle$} states. While this imaging can perfectly resolve the $L,\,R,\,B$ parts in the BZ within each spin component (see Fig.\ 3A in the main text), due to the short TOF the imaged size of these parts remains dominated by the in situ cloud size and is therefore a convolution of the quasimomentum and real-space distributions. To extract the phase after the interferometry, we count the population of atoms in the two spin states. Depending our analysis method, we count either the atoms in the individual slices (L, R, B) or all the atoms of the cloud.

\subsubsection{Acceleration parameters} 

In all the experimental runs, the magnetic field gradient produces an acceleration \smash{$\abs{\mu\nabla B}/m=2.9(1)\,$m/s$^2$}. The frequency of lattice beam three is swept via acousto-optical modulators to accelerate the atoms along the propagation direction of the beam \cite{Dahan1996}. The magnitude of this acceleration is \smash{$\abs{\vec{a}}=\frac{2}{3}\lambda_{\text{L}}\frac{d\nu}{dt}$}, where $\frac{d\nu}{dt}$ is the rate of frequency change. In the experiment, we vary the lattice acceleration from 1 to 11$\,$m/s$^2$ to change the atoms' final quasimomentum $\kyfin$. We ensure that the motion is adiabiatic for this range of forces by checking that the occupation of higher bands is negligible. 

\subsubsection{Optimization of the spin-echo sequence} 

Time-dependent fluctuations of the magnetic field $B(t)$ are the dominant source of noise in the interferometry sequence and contribute to the dynamical phase (see eq.\ \ref{eq:phidyn}). To minimize the effect of magnetic field fluctuations, which are mostly due to background AC-noise, we synchronize the beginning of the interferometer sequence to the 50\,Hz-line and keep the duration of the  sequence fixed at 1.6\,ms. 

As mentioned in Sec.\ \ref{sec:bloch_interferometry}, the orthogonality of the lattice and gradient force is crucial for the cancellation of dynamical phases, and it is necessary to fine-tune the alignment of the forces. Therefore, to create the gradient, we use both a main coil and a second fine-tuning coil which is roughly perpendicular in position to the main coil. By changing the current through the latter coil, we can tilt the direction of the combined gradient. For orthogonal gradient and lattice forces, the phase measured by the zero-area reference interferometer should be independent of $\kyfin$. Hence, to optimize the direction of the gradient, we measure the phase of the reference loop for different lattice accelerations and currents of the fine-tuning coil. From our calibration, we estimate an error on the gradient alignment of at most 2$^{\circ}$.

\end{document}